\documentclass{webofc}
\usepackage[varg]{txfonts}   % Web of Conferences font
\begin{document}
\title{Photon-triggered jets as probes of multi-stage jet modification}

\author{\firstname{Chathuranga} \lastname{Sirimanna}\inst{1,2}\fnsep\thanks{\email{chathuranga.sirimanna@duke.edu}} \and
        \firstname{Yasuki} \lastname{Tachibana}\inst{3} ~for the JETSCAPE Collaboration 
%        \firstname{Third author} \lastname{Third author}\inst{3}\fnsep\thanks{\email{Mail address for last
%             author if necessary}}
        % etc.
}

\institute{Department of Physics, Duke University, Durham, NC 27708, USA
\and
           Department of Physics and Astronomy, Wayne State University, Detroit MI 48201, USA.
\and
           Akita International University, Yuwa, Akita-city 010-1292, Japan.
          }

\abstract{%
  Prompt photons are created in the early stages of heavy ion collisions and traverse the QGP medium without any interaction. Therefore, photon-triggered jets can be used to study the jet quenching in the QGP medium. In this work, photon-triggered jets are studied through different jet and jet substructure observables for different collision systems and energies using the JETSCAPE framework. Since the multistage evolution used in the JETSCAPE framework is adequate to describe a wide range of experimental observables simultaneously using the same parameter tune, we use the same parameters tuned for jet and leading hadron studies. The same isolation criteria used in the experimental analysis are used to identify prompt photons for better comparison. For the first time, high-accuracy JETSCAPE results are compared with multi-energy LHC and RHIC measurements to better understand the deviations observed in prior studies. This study highlights the importance of multistage evolution for the simultaneous description of experimental observables through different collision systems and energies using a single parameter tune.
}
\maketitle
\section{Introduction}
\label{intro}
Heavy ion collisions offer a means to explore the early stages of the universe, particularly the Quark Gluon Plasma (QGP) formed microseconds after the Big Bang. Among the various probes employed to study QGP in heavy ion collisions, prompt photons are crucial \cite{Wang:1996yh}, as they enable the estimation of energy and direction for the initiating parton on the away side jet before energy loss occurs. In recent years, there has been a growing interest in investigating observables related to prompt photons within both theoretical and experimental communities. The experimental analysis of identifying prompt photons from a large number of shower and fragmentation photons poses significant challenges, and a common practice involves applying isolation criteria. Given that photons do not interact with the QGP medium, these isolated photons predominantly consist of prompt photons.

The scarcity of events featuring prompt photons is a consequence of the low cross-section in hard scattering processes leading to their production. In theoretical simulations, generating modified events with prompt photons requires adjusting the cross-section to significantly reduce computation time. However, it's important to note that this simulation approach differs from experimental analysis, which relies on isolated direct photons.

In this study, we compare various observables using both modified prompt photon events and full events generated by JETSCAPE 3.X (utilizing PP19\cite{JETSCAPE:2019udz} and AA22\cite{JETSCAPE:2022jer} tunes) against available experimental results. Additionally, for the first time, we explore isolated photon and di-jet correlation, groomed jet substructure associated with $\gamma$-triggered jets, and $\gamma$-jet asymmetry, using new JETSCAPE results featuring improved statistics \cite{Sirimanna:2020ggn}.

\section{Simulating photon triggered jets using JETSCAPE framework }
\label{sec-JSsimulation}

JETSCAPE is the first framework that supports multistage evolution, where multiple modules can be employed to simulate the partonic shower within a QGP medium, depending on the virtuality of each shower parton. This multistage evolution capability enables the JETSCAPE framework to simultaneously describe numerous experimental observables across different center-of-mass energies and centralities using a single parameter tune.

In the simulation of p-p collisions, partons produced in the initial hard scattering in PYTHIA are fed into the MATTER energy loss module for vacuum partonic showers. Subsequently, all final state partons from MATTER are transferred back to PYTHIA for string hadronization. For Pb-Pb collisions, pre-generated event-by-event hydro profiles with initial conditions are utilized. Similar to p-p collisions, partons produced in the initial hard scattering in PYTHIA are transferred into MATTER and LBT energy loss modules based on their virtuality. Final state hadrons are generated using PYTHIA's string hadronization, following the partonic shower. The same set of parameters used in previous p-p\cite{JETSCAPE:2019udz} and Pb-Pb\cite{JETSCAPE:2022jer} studies is applied in this study without further parameter tuning.

In this study, trigger photons are identified using the same isolation criterion employed in experimental analysis for both modified prompt-photon events and full events \cite{ATLAS:2018dgb, CMS:2017ehl}. The same sets of generated events at $\sqrt{S_{NN}}=5.02~\text{TeV}$ are used to analyze a wide variety of observables and compare them with available experimental results. After identifying the isolated photon in an event, jets are clustered using the anti-$k_T$ jet clustering algorithm within FastJet.

\section{Results and Discussion}
\label{sec-results}

\begin{figure}[ht]
\centering
\includegraphics[width=\textwidth,clip]{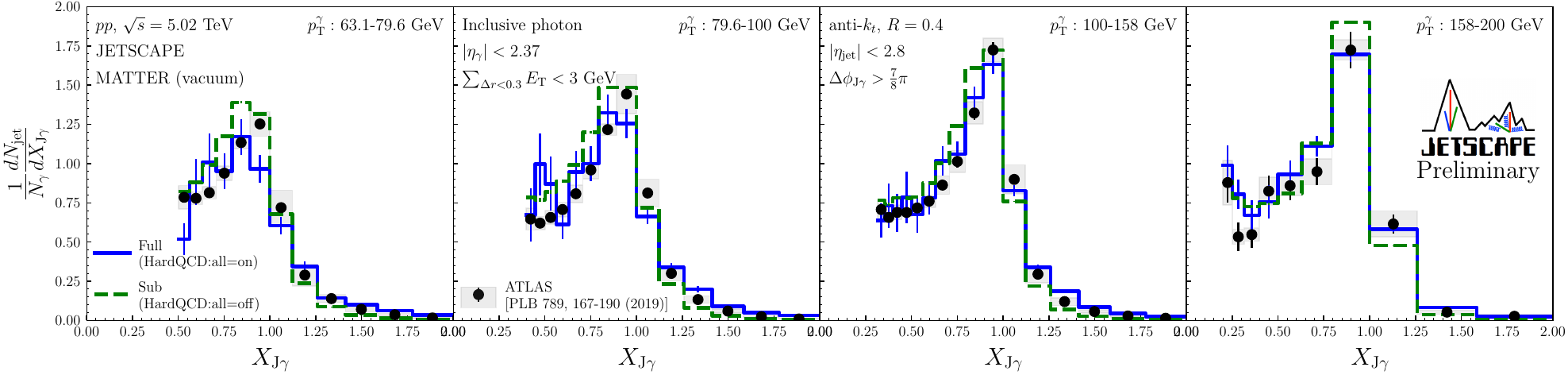}
\caption{$\gamma$-jet asymmetry for p-p collisions using prompt-photon events and full events generated by JETSCAPE compared with ATLAS results. Here four $p_{T\gamma}$ regions from $63.1 GeV$ to $200 GeV$ are used. }
\label{fig-1}       % Give a unique label
\end{figure}

\begin{figure}
\centering
\sidecaption
\includegraphics[width=\textwidth,clip]{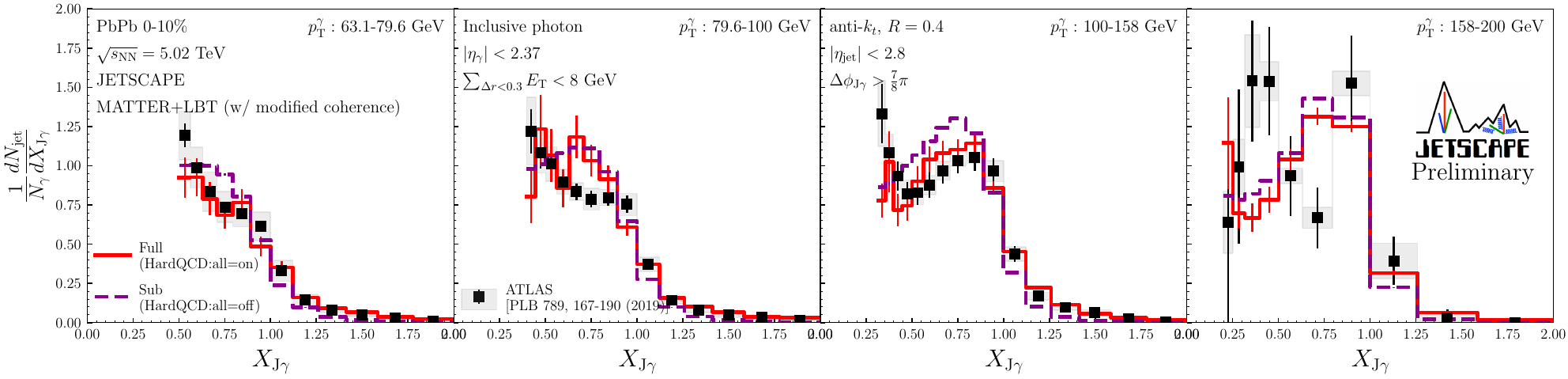}
\caption{$\gamma$-jet asymmetry for Pb-Pb collisions using prompt-photon events and full events generated by JETSCAPE compared with ATLAS results. Same $p_{T\gamma}$ regions as Figure \ref{fig-1} is used here.}
\label{fig-2}       % Give a unique label
\end{figure}

The $\gamma$-jet asymmetry was investigated for both p-p and Pb-Pb collisions using modified prompt-photon and full events, and the findings were compared with those of ATLAS and CMS. Figures \ref{fig-1} and \ref{fig-2} illustrate comparisons of $X_{j\gamma}$ distributions for four different $p_T^{\gamma}$ intervals with ATLAS results for p-p and Pb-Pb, respectively. Given that ATLAS results are unfolded, unmodified JETSCAPE results were used in this comparison. Despite the considerably large uncertainties in the results from full events, due to the rarity of events with prompt photons, a better agreement can be observed in both p-p and Pb-Pb results.  

\begin{figure}
\centering
\sidecaption
\includegraphics[width=0.6\textwidth,clip]{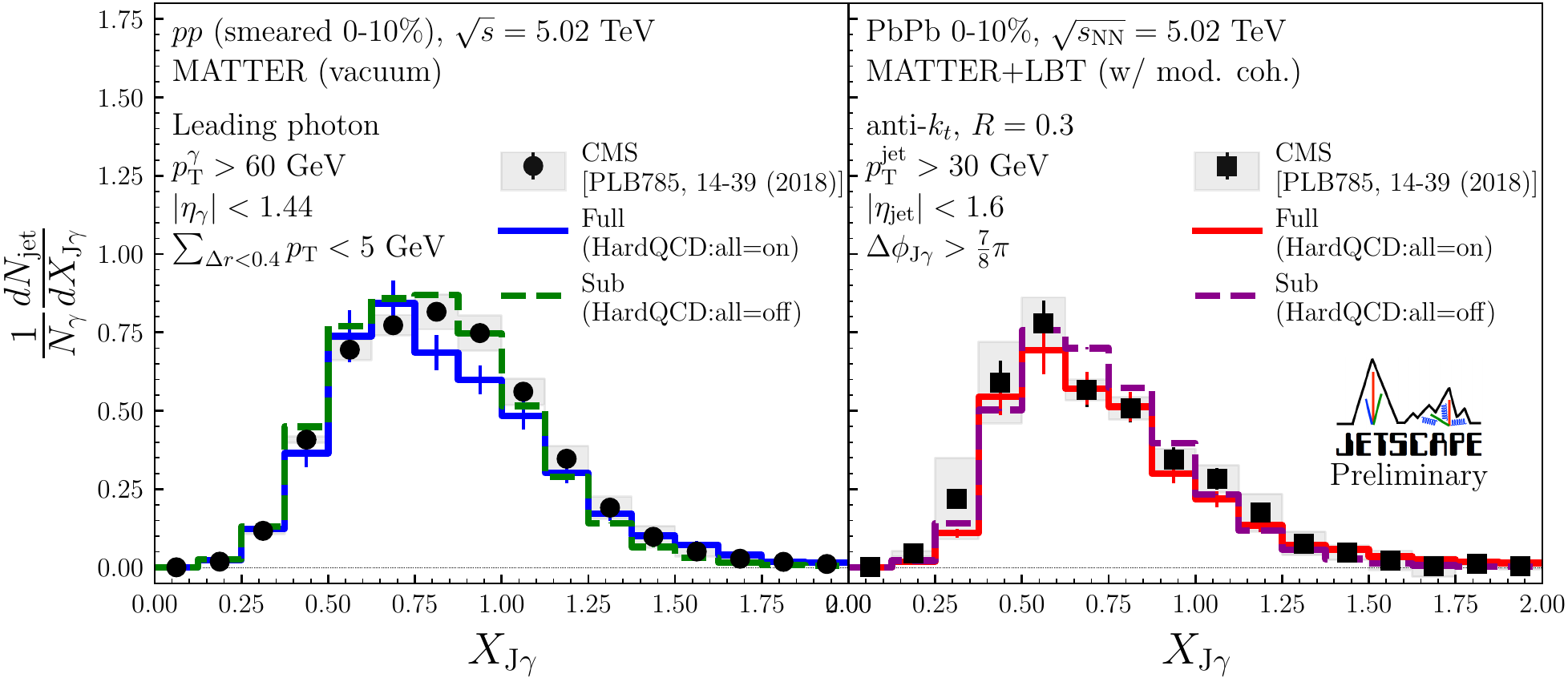}
\caption{$\gamma$-jet asymmetry for p-p and Pb-Pb (0-10\%) collisions using prompt-photon events and full events generated by JETSCAPE compared with CMS results. Here both p-p and Pb-Pb distributions are smeared according to the smearing function for 0-10\% centrality.}
\label{fig-3}       % Give a unique label
\end{figure}

Since CMS results for both p-p and Pb-Pb are smeared, the same smearing function is applied for a proper comparison, as illustrated in Figure \ref{fig-3}. Similarly, in this case, results from full events exhibit better agreement with the experimental results. Although the isolated photons mainly consist of prompt photons, these findings suggest a significant contribution to the transverse momentum imbalance from other photons, including those produced in the partonic shower and fragmentation photons.

\begin{figure}
\centering
\sidecaption
\includegraphics[width=\textwidth,clip]{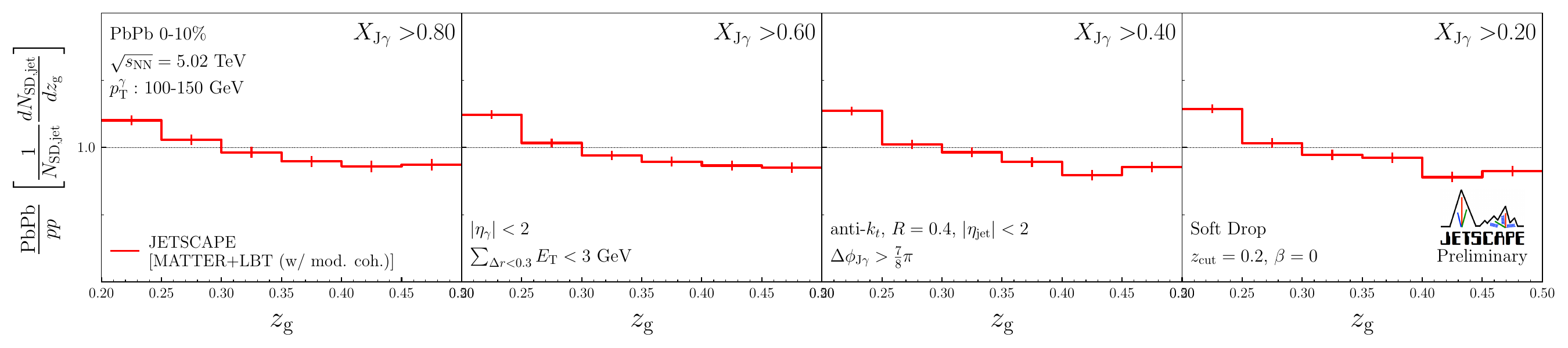}
\caption{$z_g$ distribution using photon-triggered jets calculated for four different $X_{J\gamma}$ regions by using prompt-photon events generated by JETSCAPE.}
\label{fig-4}       % Give a unique label
\end{figure}

Figure \ref{fig-4} displays the $z_g$ distribution, which measures the energy imbalance of the hardest split in a $\gamma$-triggered jet. The ratio between Pb-Pb and p-p illustrates the modification from the QGP medium. As observed in Figure \ref{fig-4}, the $z_g$ distribution does not exhibit a significant dependence on the transverse momentum imbalance, $X_{j\gamma}$. Although there are no experimental results available yet to compare with these groomed jet substructure observables using photon-triggered jets, ongoing experimental studies are being conducted. 

\begin{figure}
\centering
\sidecaption
\includegraphics[width=0.7\textwidth,clip]{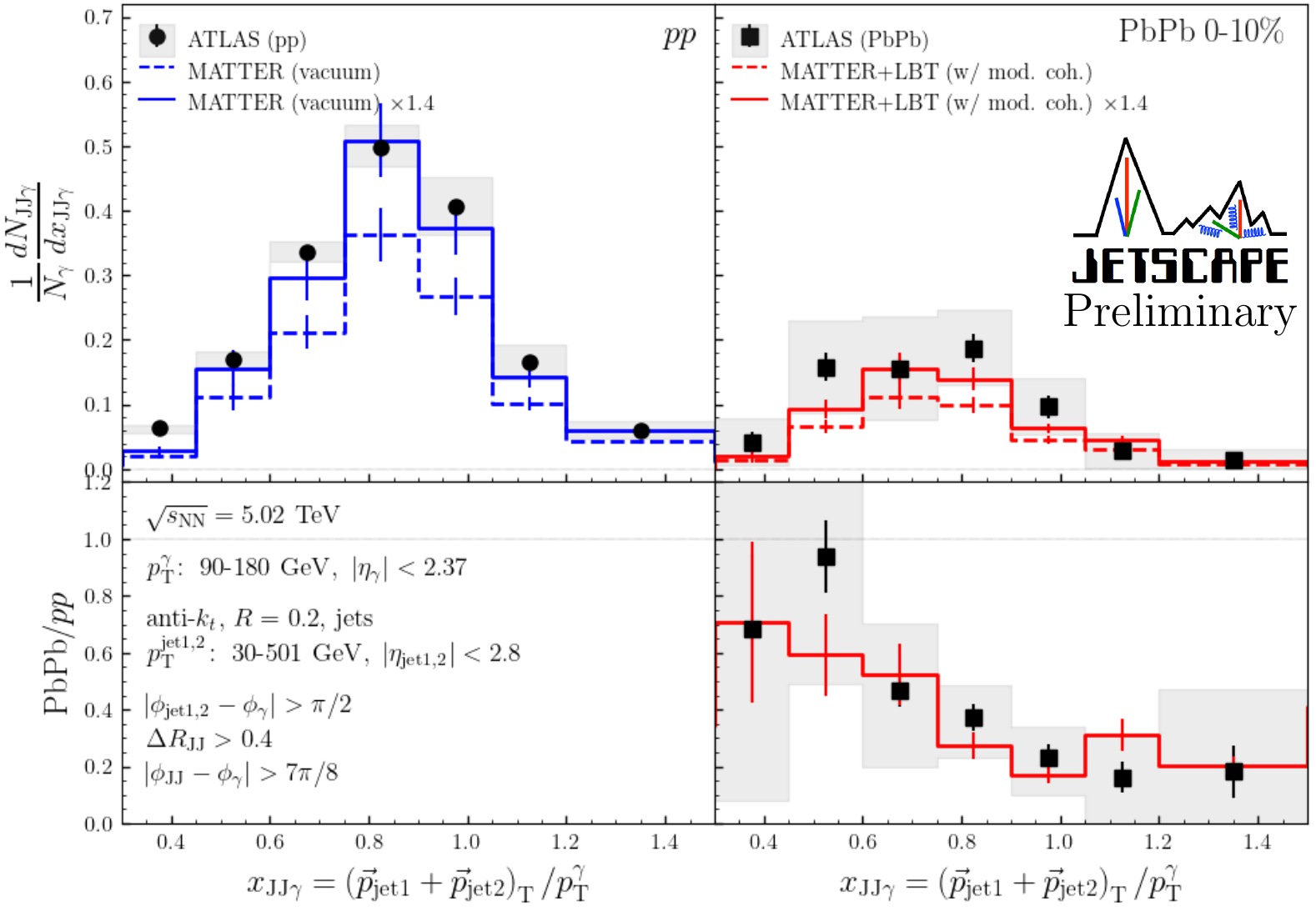}
\caption{Isolated photon and di-jet correlation, $x_{JJ\gamma}$ calculated using prompt-photon events generated by JETSCAPE compared with ATLAS preliminary results \cite{ATLAS-CONF-2023-008}. }
\label{fig-5}       % Give a unique label
\end{figure}

The isolated photon and multi-jet correlation is another recent experimental measurement by ATLAS collaboration\cite{ATLAS-CONF-2023-008}. In Figure \ref{fig-5}, the $x_{JJ\gamma} = (\vec{p}_1 + \vec{p}_2)_T / p_{T,\gamma}$ distribution using JETSCAPE for p-p, Pb-Pb, and the ratio between them is compared to ATLAS preliminary data. Although JETSCAPE shows excellent agreement with the ratio, a multiplication factor of 1.4 is required to accurately describe both p-p and Pb-Pb results. Since no model can explain these results for isolated photon and di-jet correlation without utilizing such a multiplication factor, additional theoretical studies are necessary to comprehend these observables. 

JETSCAPE results for the majority of photon-triggered jet observables provide an excellent description of the experimental results without requiring additional parameter tuning. While current theoretical knowledge is insufficient to fully understand certain observables like isolated photon and multi-jet correlation, JETSCAPE, with its multistage evolution, offers an excellent description of all stages of jet evolution — a significant improvement over single-stage jet evolution. This analysis was conducted without further modification of the previous PP19 and AA22 parameter tunes. Therefore, this study serves as a parameter-free verification of multistage evolution. 

\section*{Acknowledgments}

These proceedings are supported in part by the National Science Foundation (NSF) within the framework of the JETSCAPE collaboration, under grant numbers ACI-1550300 OAC-2004571 (CSSI:X-SCAPE) and in part by the U.S. Department of Energy (DOE) under grant number DE-SC0013460.

%
% BibTeX or Biber users please use (the style is already called in the class, ensure that the "woc.bst" style is in your local directory)
\bibliography{PhotonTriggeredJets}

\end{document}